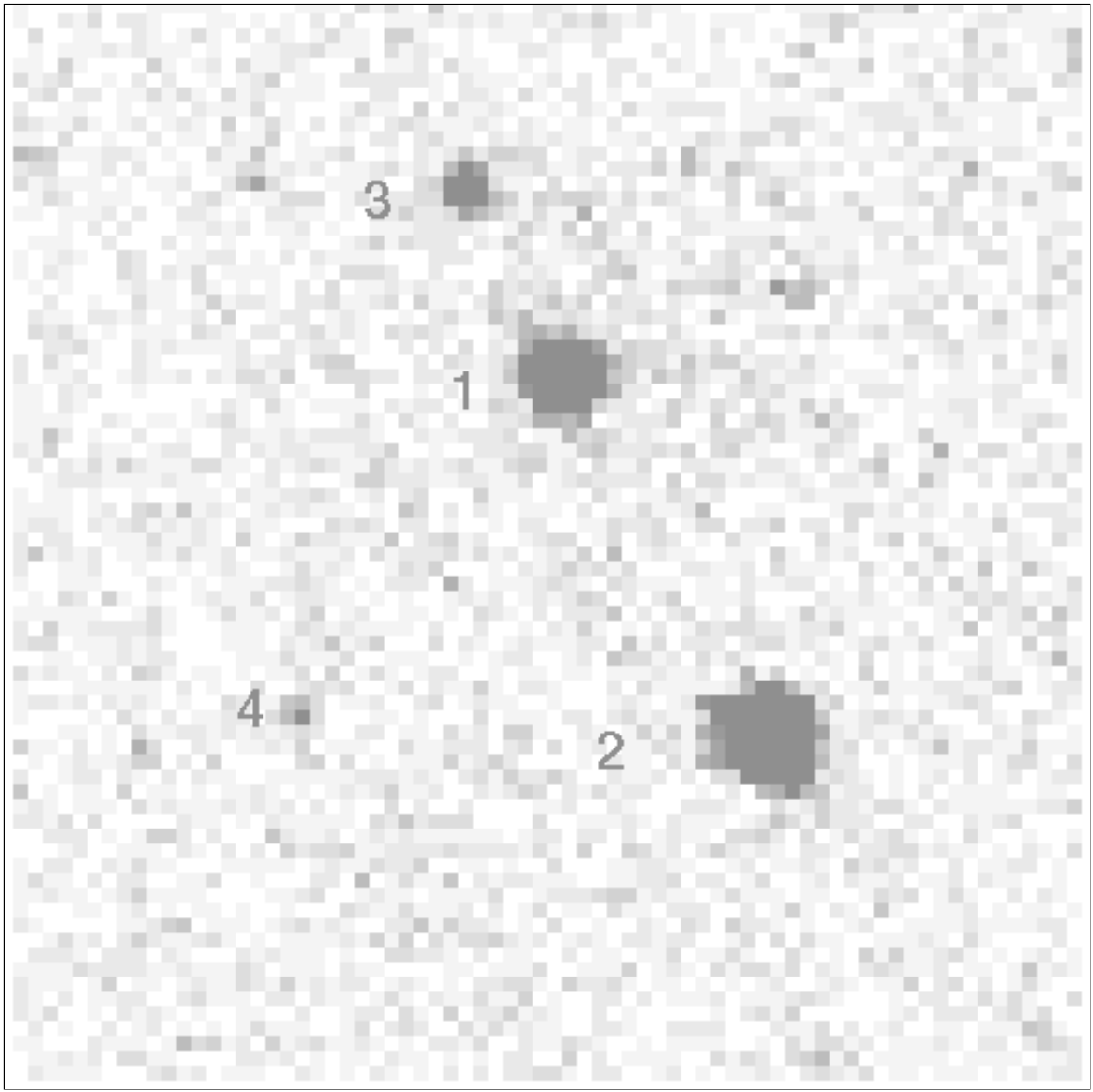

Frame       : lsi
Identifier  :
ITT-table   : ramp.itt
Coordinates :        ,      :      ,
Pixels      : 1, 1 : 512, 512
Cut values  : 0, 10
User        : goldoni

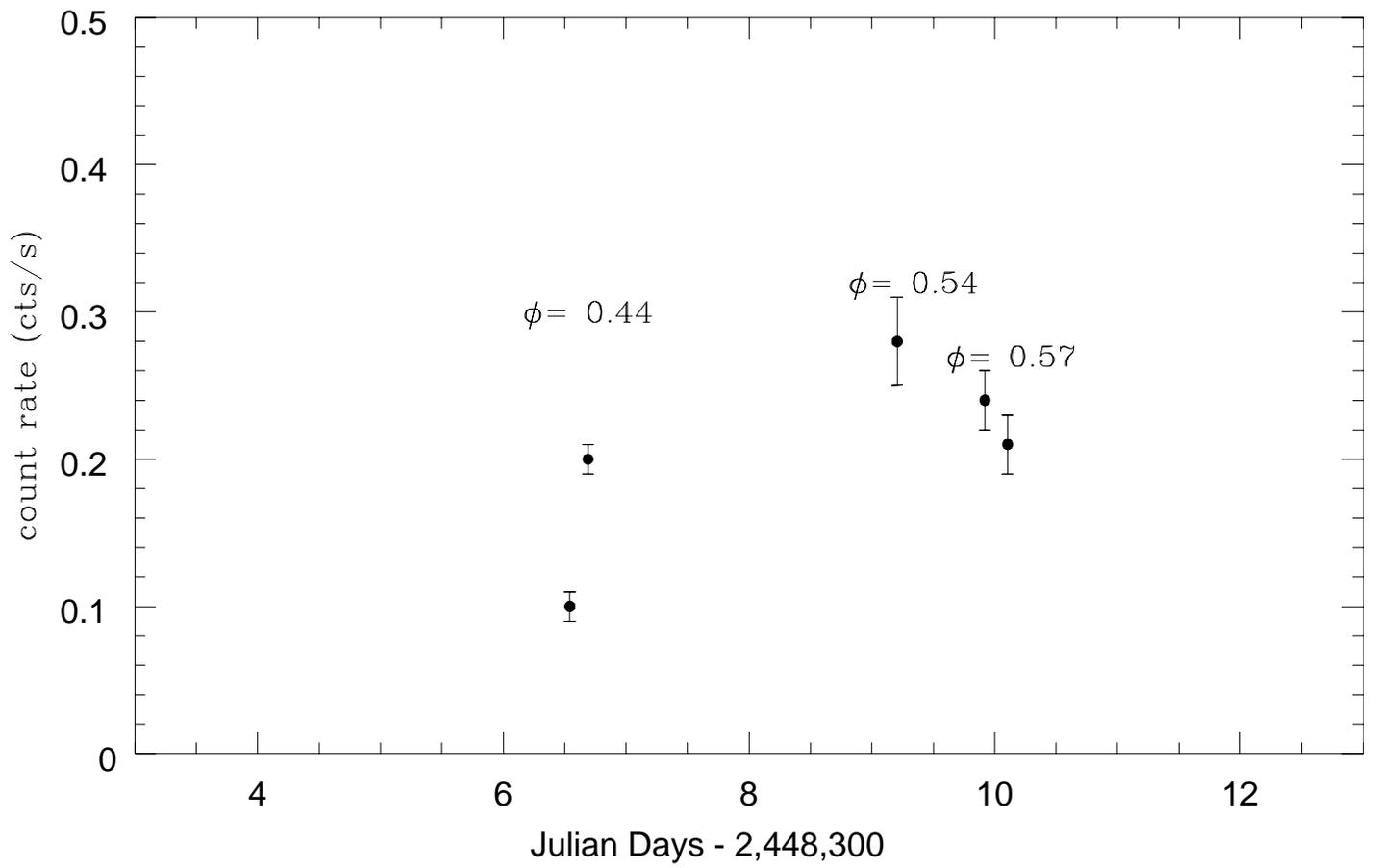

# X–ray observations of the peculiar Be star LSI +61° 303

**Paolo Goldoni**[1,2] and **Sandro Mereghetti**[1]

[1] Istituto di Fisica Cosmica, CNR, v. Bassini 15, I-20133, Milano, Italy; e-mail: sandro@ifctr.mi.cnr.it
[2] Dipartimento di Fisica, Università di Milano, v. Celoria 16, I-20133, Milano, Italy; e-mail: goldoni@ifctr.mi.cnr.it

**Abstract.**

We present the results of a ROSAT Position Sensitive Proportional Counter observation of the variable radio star LS I +61° 303. The source has a power law spectrum with photon index $1.1 \pm 0.3$ and a luminosity of $5.9 \times 10^{33}$ erg s$^{-1}$ in the 0.1 -2.4 keV range. Variability on a timescale of days is clearly visible in the ROSAT data, but no periodic pulsations have been detected. The low X–ray luminosity is difficult to explain in the context of the models involving super Eddington accretion which have been proposed to account for the non thermal radio outbursts. Though the weak X–ray flux is marginally compatible with the levels expected from normal B and Be star winds, the variability and the rather hard spectrum are more suggestive of non thermal emission. This might originate in a shocked region produced by a young pulsar in orbit around the Be star.

## 1. Introduction

The early type star LS I +61° 303 is the optical counterpart of GT0236+610 (Gregory & Taylor 1978), a variable radio source characterized by strong, non thermal outbursts occurring with a periodicity of 26.5 days (Taylor & Gregory 1982, 1984). The peak flux density during the outbursts can reach the value of $\sim 300$ mJy (Taylor et al. 1992), which, for an assumed distance of 2.3 kpc (Gregory et al. 1979), corresponds to a 10 GHz luminosity of $\sim 2 \times 10^{31}$ erg s$^{-1}$. The radio outbursts are of variable amplitude and shape, and there is some evidence that their intensity varies with a period of 4 years (Gregory et al. 1989; Paredes et al. 1990). High resolution radio maps obtained with the VLBI technique (Massi et al. 1993) show that the radio outbursts are produced by synchrotron emission from a two milliarcsecond double source expanding at $\sim 5 \times 10^7$ cm s$^{-1}$.

Optical observations showed the presence of strong and broad hydrogen emission lines, and led to a classification of LS I +61° 303 as a main sequence B0-1 star (Gregory et al. 1979). Radial velocity measurements, though compatible with a low mass ($\sim 1 M_\odot$) secondary in an eccentric orbit with the 26.5 days period derived from the radio observations, were not conclusive owing to the intrinsic variations of the spectral features (Hutchings & Crampton 1981). A faint modulation at the radio period has also been detected in the optical ( Paredes et al. 1994, Mendelson & Mazeh 1994) while the presence of a periodicity in the near infrared is controversial (Paredes et al. 1994, Hunt et al. 1994).

Since the time of its discovery, this object attracted considerable interest, being located inside the error region of a strong COS-B $\gamma$-ray source (Bignami & Hermsen 1983). Recent observations with EGRET (Fichtel et al. 1994), which have confirmed the presence of emission at $E > 100$ MeV and reduced the error box to a circle of 33 arcmin radius, support the possible association of LS I +61° 303 with the $\gamma$-ray source.

The models that have been proposed to explain the radio (and possible high energy) emission from the star LS I +61° 303 are based on two different scenarios, both involving an eccentric binary system with orbital period equal to 26.5 days: supercritical accretion onto a compact companion (Taylor & Gregory 1982,1984, Taylor et al. 1992) or non-thermal processes related to the presence of a young, fastly spinning neutron star (Maraschi & Treves 1981, Lipunov & Nazin 1994).

The radio properties of LS I +61° 303 are similar to those of peculiar X–ray binaries containing compact objects, like Cyg X–3, Cir X–1 and SS 433 (Hjellming & Penninx 1991). However, there is no direct evidence (e.g. X–ray pulsations or bursts) for the presence of an accreting neutron star in this source. In particular, it is remarkable that the X-ray luminosity of LS I +61° 303 is only of the order of $L_x \sim 10^{33}$ erg s$^{-1}$, i.e. 2 to 5 orders of magnitude lower than in the above sources. The weak X–ray flux ($\sim 4 \times 10^{-12}$ erg cm$^{-2}$s$^{-1}$) prevented detailed studies of LS I +61° 303 with collimated instruments such as EXOSAT or GINGA, and most of our knowledge on its X–ray properties is based on Einstein observations (Bignami et al. 1981). These short observations, carried out about fifteen years ago, did not allow to determine the spectral parameters of the X–ray emission, owing to the limited statistics ($\sim 220$ photons in $\sim 2500$ seconds) and the poor energy resolution ($\sim 50\%$) of the Einstein IPC. Here we report the results of a recent observation of LS I +61° 303 with the position sensitive proportional counter (PSPC) instrument on board the ROSAT satellite.

**Table 1.** Sources in the field

| Number | name | $m_V$ | Sp.Type | counts s$^{-1}$ | $f_x/f_v$ | notes |
|--------|------|-------|---------|-----------------|-----------|-------|
| 1 | LSI+61°303 | 10.8 | $B0-1$ | 0.22 | $2.4 \times 10^{-2}$ | Variable radio source GT0236 + 610 |
| 2 | BD 60536 | 10.2 | $G2$ | 0.21 | $1.3 \times 10^{-2}$ | Variable, also detected with EINSTEIN HRI |
| 3 | HD 16429 | 7.7 | $O9.5I$ | $1.4 \times 10^{-2}$ | $5.3 \times 10^{-5}$ | |
| 4 | BD 60544 | 9.7 | $B2$ | $2.23 \times 10^{-3}$ | $6.5 \times 10^{-5}$ | |

## 2. Observations and data analysis

The observation was performed from 1991 February 18 to 22 (JD 2,448,306.5 to JD 2,448,310), and it consists of 14 intervals with durations ranging from $\sim 500$ s to $\sim 2000$ s each, for a total net exposure time of $\sim 17000$ s. Four point sources were clearly detected near the center of the field of view (see Figure 1), and identified with the stars listed in Table 1. The central source (n.1) is, within the statistical and systematic uncertainties of the PSPC positional errors, coincident with LS I +61° 303. Using the optical positions of the other three sources, as derived from the HST GSC, we have computed the average coordinate shift to correct for the systematic PSPC positional error. The resulting position for the source n.1 is R.A.=2h 40m 31.6s, DEC.=61°13' 44"(J2000), only 2" from the optical and radio coordinates of LS I +61° 303 (Gregory et al. 1979).

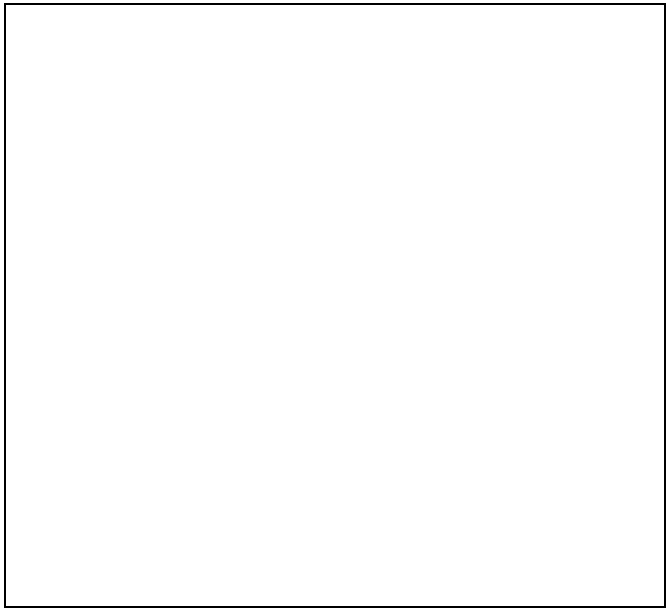

**Fig. 1.** Image of the central $\sim$18x18 arcmin$^2$ of the ROSAT PSPC field

Figure 2 shows the background subtracted light curve of LS I +61° 303. A flux increase of a factor $\sim 3$ over a timescale of $\sim 3$ days is clearly visible. An examination of the source hardness ratios does not provide any evidence for spectral variations associated with the flux increase. We have therefore performed a spectral analysis on the whole observation. The source counts were extracted from a circle of 50" radius centered on the source position and the background from a circular annulus around the four central sources. This resulted in $\sim 3800$ net counts, which were rebinned into 16 bins to achieve a minimum signal to noise ratio of 15 in each spectral bin, and then fit to different spectral models using the XSPEC software. An acceptable fit (reduced $\chi^2 = 1.37$ with 13 d.o.f.) was obtained with a power law spectrum, yielding a photon index of 1.1 $\pm 0.3$ and an absorption $N_H = 4.7(\pm 0.8) \times 10^{21}$ cm$^{-2}$. The corresponding X-ray flux in the 0.1-2.4 keV range is $4.32(\pm 2.1) \times 10^{-12}$ erg cm$^{-2}$ s$^{-1}$, which for a distance of 2.3 kpc corresponds to an unabsorbed luminosity $L_x = 5.9(\pm 2.9) \times 10^{33}$ erg s$^{-1}$. The spectral parameters derived with thermal bremmstrahlung models were less well constrained. In fact, acceptable fits were obtained for any temperature greater than $\sim 5$ keV and for $N_H = 6.0(\pm 0.3) \times 10^{21}$. According to the average relation between $N_H$ and $A_V$ (Gorenstein 1975), the column density derived with the above fits is consistent with the $\sim 3$ magnitudes of visual extinction toward LS I +61° 303 (Gregory et al. 1979). To compare with X-ray observations of isolated stars, we also considered thermal plasma emission by fitting Raymond-Smith models (Raymond & Smith 1977) with solar abundances, but these models gave unacceptable fits.

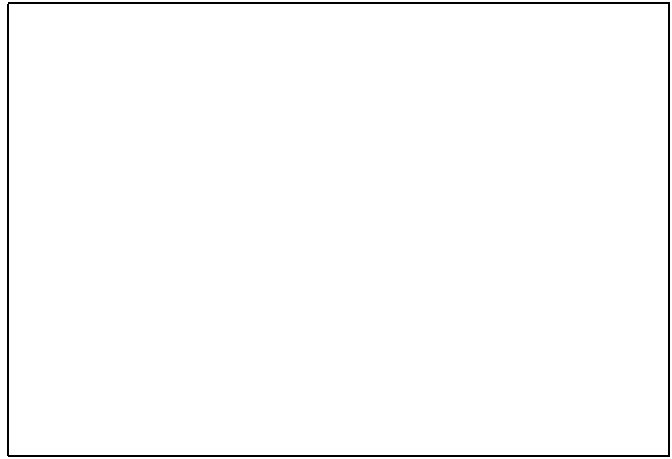

**Fig. 2.** X-ray light curve of LS I +61° 303 with 1$\sigma$ error bars

A search for periodicities was carried out using the Rayleigh test (Mardia 1972, Brazier 1994). The arrival times of the 4014 counts used in the analysis were corrected to the solar system barycenter using the IRAF/PROS software. We searched for $\sim 900,000$ statistically independent frequencies in the 0.001-10 s range, without finding any significant periodicity. In the hypothesis of a sinusoidal pulse shape, we derived a 3$\sigma$ upper limit to the pulsed fraction of $\sim 12$ %. In order to reduce pos-sible pulse smearing effects resulting from an orbital motion, the search was also repeated for the three separated observing periods, but this yielded again a negative result ($3\sigma$ upper limits of the order of 20 %).

## 3. Discussion

The soft X–ray luminosity of LS I +61° 303 is so low, compared to that of typical High Mass X–ray Binaries, that one should first consider the possibility that it result from wind emission in the B star itself. The bolometric luminosity of $1.5 \times 10^{38}$ erg s$^{-1}$ (Gregory et al. 1979), yields for LS I +61° 303 a value of $(L_x/L_{bol}) \sim 4 \times 10^{-5}$ which, though higher than the average, is compatible with the range of values found in ROSAT observations of B and Be stars (Meurs et al. 1992). However, the spectrum derived from our data is rather hard, compared to the typical thermal spectra with kT < 1 keV observed in B stars (Cassinelli et al. 1994).

Indeed the hardness ratios for LS I +61° 303, 0.99 ±0.02 (0.4-2.4 keV to 0.1-2.4 keV) and 0.81 ± 0.02 (1.0-2.4 keV to the 0.4-2.4 keV), differ from those of most of the O and B stars (with or without emission lines) detected by Meurs et al.. On the other hand, they are similar to those of the Be star X Per, which likely contains an accreting neutron star (White et al. 1977), and of the PSR 1259-63/SS2883 system (Cominsky et al. 1994). The latter source is the first radio pulsar discovered with a Be star companion (Johnston et al. 1992, 1994). Its weak X–ray luminosity (Cominsky et al. 1994) is very unlikely to result from accretion (Campana et al. 1994), while it can be explained by shock emission where the relativistic pulsar wind collides with the mass outflowing from the Be companion (Tavani et al. 1994). Our ROSAT results for LS I +61° 303, i.e. the hard power law spectrum, the variability and the absence of pulsations, are compatible with a similar non–thermal mechanism, as first proposed by Maraschi & Treves (1981).

Alternatively, if LS I +61° 303 contains a neutron star accreting from the equatorial component (Waters et al. 1988) of the Be star's wind, its luminosity should be at least three orders of magnitude higher than the observed value.

If the orbit of the accreting object lies within the equatorial wind component, Taylor et al. (1992) have shown that a moderate eccentricity ($e > 0.4$) is enough to cause supercritical accretion near the periastron. The range of orbital phases in which supercritical accretion occurs (about 10 days centered slightly after periastron) does not change significantly for higher eccentricity values. Our data were taken at phases $\phi \sim 0.44$, $\phi \sim 0.54$, and $\phi \sim 0.57$ of the Taylor & Gregory (1982) ephemerides, in which periastron is at $\phi \sim 0.2$. The radio outburst generally occurs between phases $\sim 0.4$ and $\sim 0.8$, peaking at $\phi \sim 0.6$ (Taylor et al.1992). The X–ray data were thus taken during a period of expected high accretion, difficult to reconcile with the low observed flux. Though the exact time of periastron passage is not very precisely constrained, we note that the model of Taylor et al. (1992) predicts an accretion of the order of $\dot{M}_{acc} \sim 2 \times 10^{-9}$ $M_\odot$ yr$^{-1}$ even at apoastron. A possible solution is to assume that the bulk of the accretion luminosity is emitted in $\gamma$-rays rather than in the classical X–ray range. This would be a unique case in the domain of compact objects accreting in binary systems. In this respect we note that the possible detection in the 1-30 MeV range with COMPTEL/GRO (van Dijk et al. 1994) corresponds to a luminosity of $\sim 10^{36}$ erg s$^{-1}$.

Another possibility for reducing the X–ray efficiency of the accretion process is to assume that the matter flow is stopped at the magnetospheric radius by the centrifugal barrier (Stella et al. 1994). As noted by Campana et al. (1994), this situation can occur in LS I +61° 303 for reasonable values of the neutron star magnetic field and spin period (e.g. $P \lesssim 10$ $s$ for $B \sim 10^{12}G$ or $P \lesssim 2$ $s$ for $B \sim 10^{11}G$). An attractive feature of this hypothesis is that it naturally explains the absence of X–ray pulsations, since the accretion flow cannot proceed down to the neutron star along the polar field lines.